# A Survey on the State of the Practice in Distributed Software Development: Criteria for Task Allocation

Ansgar Lamersdorf
University of Kaiserslautern
Kaiserslautern, Germany
a_lamers@informatik.uni-kl.de

Jürgen Münch
Fraunhofer IESE
Kaiserslautern, Germany
Juergen.Muench@iese.fraunhofer.de

Dieter Rombach
University of Kaiserslautern and Fraunhofer IESE
Kaiserslautern, Germany
Dieter.Rombach@iese.fraunhofer.de

*Abstract*—The allocation of tasks can be seen as a success-critical management activity in distributed development projects. However, such task allocation is still one of the major challenges in global software development due to an insufficient understanding of the criteria that influence task allocation decisions. This article presents a qualitative study aimed at identifying and understanding such criteria that are used in practice. Based on interviews with managers from selected software development organizations, criteria currently applied in industry are identified. One important result is, for instance, that the sourcing strategy and the type of software to be developed have a significant effect on the applied criteria. The article presents the goals, design, and results of the study as well as an overview of related and future work.

*Key words: Global Software Development, task allocation, project management*

## I. INTRODUCTION

As global and distributed software development (GSD) is "becoming a norm in the software industry" [3], effective strategies for organizing it have become critical. An important decision for organizing distributed work is the allocation of tasks over the distributed sites: Task distribution can influence both the benefits of global software development (e.g., cost reduction, availability of people, proximity to the customer) and its risks (e.g., inexperienced workforces, communication overhead).

There are several possible strategies for allocating tasks that use different criteria for allocation. Example criteria are:
- Costs: Bass and Paulish state that in practice, distributing work to low-cost countries has become a cost-saving strategy for many organizations [1].
- Time zone differences: Conflicting strategies exist that use time zone differences as criteria [2]: One strategy suggests assigning work within "time-zone bands" in order to allow for synchronous communication. Another strategy suggests assigning work to different time zones in order to reduce development time by using a "follow-the-sun" model.
- Cultural differences: There are also different strategies that use cultural differences as criteria for work allocation: One strategy is to assign work to regions close to the customer in order to reduce differences between developing site and customer site [4]. Another strategy aims at minimizing the differences between development sites by assigning work to people located in different cultures [9].

All of these criteria may have an impact on the success of a specific task distribution. Thus, besides the project goals and project characteristics, a systematic task allocation needs to consider many different criteria. A first step towards supporting task distribution decisions can be seen in an analysis of criteria applied in practice. This article presents an analysis of criteria that are currently applied in distributed development projects. The results are based on an interview study with practitioners in distributed software development. By conducting interviews with experts of many different environments, we tried to get an overview on the current state of the practice in task allocation. Even though the number of interviewees is rather limited, the study results in a classification of GSD types that to our belief includes most of the distributed software development practices currently applied.

The article is structured as follows: First, the related work with respect to identifying criteria for task allocation is presented. After that, the qualitative study is explained in detail. The results of the study are presented in section four, followed by a summary and future work.

## II. RELATED WORK

The following section is divided into empirical studies that were performed for analyzing task allocation practices and theoretical models that define their own criteria for task allocation.

### A. Empirical Studies of Work Organization and Task Allocation

Several studies have been performed in order to identify the practices of task allocation in distributed development. Similar to this study, the information was usually gathered by interviewing practitioners.

Grinter, Herbsleb, and Perry [7] examined software development projects at Lucent Technologies with the goal of analyzing the coordination of work. They identified four different ways of how work is organized: 1) *Functional area of expertise*: Specific functional expertise is located at single sites. Task allocation is thus done according to specialization and expertise. 2) *Product structure*: The organization is split up by the architecture of the product to be developed. 3) *Process steps*: The work is broken into process steps that are assigned to the different sites. 4) *Customization*: One site owns and develops the core code for the product, while customer- or market-specific alterations are assigned to other sites that are close to the specific markets.

The study focuses primarily on general types of task allocation. For two of the types (functional area of expertise, customization), criteria for task allocation are named (specific expertise at the sites, proximity to markets). A more comprehensive set of criteria for task allocation, however, is not identified.

Similar types of work allocation are described by Mockus and Weiss [11]. They distinguish between work transfer by functionality, localization, development stage, and maintenance stage. But, again, criteria for task allocation are not listed explicitly.

A study by Edwards et al. [5] also looked at the organization of work and the decomposition of projects. Contrary to the other publications, Edwards et al. additionally identified criteria for task assignment that were applied in practice. However, they looked at task assignment to individuals rather than assignment to sites (as done in this article). In their study, individual expertise was the most popular criterion for task assignment. Minor criteria were availability and process ownership. The study assumes that task allocation is driven by only one single criterion in every project. It thus identified only a small number of criteria. In the study presented in this article, we assume that many, potentially interdependent, different criteria are considered in task allocation, which results in a more detailed list of criteria.

Westner and Strahringer [20] performed a very similar study to the one presented in this article, but from a different perspective: Instead of looking for criteria for assigning tasks of a given project to sites, their study aimed at identifying criteria for selecting projects for which offshored developed was an option. The criteria identified in interviews were compared to a list of criteria gathered from a preceding literature study. As a result, a detailed list of criteria was found, with the size and the documentation of the project being the most important ones.

### B. Theoretical Models

There exists a small number of models for decision support in work allocation that come with a set of criteria.

One model by Mockus and Weiss [11] uses two criteria for the decision: minimization of the number of tasks spanning multiple sites and availability of free resources per site. These criteria are based on previous empirical studies. However, other criteria that could be relevant for task allocation in specific contexts are not considered.

Setamanit et al. [18] defined a simulation model for evaluating task allocation strategies. It simulates the effects of distributed development on project duration. The results are dependent on a set of input variables that describe the available sites (e.g., productivity, development quality) and the relationship between the sites (e.g., distance, cultural difference, familiarity). Thus, these variables represent criteria that influence the success of task allocation. However, they are not gathered empirically.

Further models are, for instance, presented in [19], [15].

Other criteria for task allocation are given in offshore attractiveness models. These models, often developed by consulting companies, weigh geographical regions in order to give decision support for offshoring work.

One model is given by A.T. Kearney [8]. It is not focused on software development but on remote services in general, which include IT services and software development. The model uses so-called "index metrics" as criteria. These include costs for wages and taxes, skills and availability of people, and the general business environment. A similar model is presented in [13]

In general, the theoretical models deliver very different sets of criteria for evaluating or selecting task allocation strategies. However, these criteria focus only on selected aspects and have mostly not been gathered empirically. Offshore attractiveness models provide criteria that are usually not focused on software development but on offshoring in general.

## III. THE QUALITATIVE STUDY

The following section will give an overview of the terminology and the goal of this study, its design, and its execution.

*A. Terminology and Study Goal*

When software development is distributed across a number of sites, every task within the development must be assigned to one or more sites. This is defined as *task allocation*. Task allocation is a decision made during project management.

We assume that every task allocation follows a set of *criteria*. A criterion for task allocation directs the decision according to a certain characteristic (e.g., "labor cost"). Usually, task allocation follows different, sometimes conflicting criteria that are (implicitly) weighted.

The goal of this study can be described as follows: From the viewpoint of a project manager in a GSD project, the practice of task allocation will be analyzed at different GSD organizations in order to identify the criteria used for task allocation. From that goal, the following research questions are derived:

*Question 1*: What is the organization's general background in global software development?

*Question 2*: Which criteria are applied for task allocation in GSD projects?

*B. Design*

The research questions were investigated using qualitative interviews with practitioners.

1) *Case and Sampling*: For the interviews, we looked for subjects with experience in distributed software development from a project management perspective. Most of the interviewees were selected via personal contacts of the authors. Others were selected by contacting participants of the 2008 International Conference on Global Software Engineering. The interviewees were thus not randomly selected, which is, for instance, reflected in the fact that a relatively large proportion (four interviewees) came from a background of developing software for US governmental agencies.

Twelve practitioners from eleven companies were interviewed. Most of the participants had experiences in distributed software development in middle or senior management positions (e.g., project, quality, or product manager). Other positions included chief architect or process analyst. The interviewed persons had many years of experience in distributed development – the majority reported from at least five years experience and two of them had been involved in distributed development for nearly 20 years.

The participants' companies came from many different application domains, such as satellite development, educational software, and software services. All of them were medium-sized or large with the smallest company having about 900 employees. Out of the ten companies, eight were US-based, one was based in Europe, and was one based in India.

2) *Instrumentation*: The study was conducted in four steps. The research methodology mainly followed the guidelines given in [17].

*1. Literature study*: A prior literature study was performed in order to get a general overview of possible criteria for task allocation. The literature study analyzed 26 conference and journal publications and resulted in 15 possible criteria. It is described in detail in [10].

*2. Formulation of the questionnaire*: A standardized questionnaire was used for the interviews. The questionnaire was designed by refining the study goal and research questions into more specific questions.

*3. Interview conduction*: As common in qualitative studies [17], the interview was semi-structured with a mixture of closed and open questions. Interviews were conducted in person or over the telephone. Telephone interviews have some drawbacks compared to interviews in person [12], especially with complex questions, and there is a higher chance of misunderstandings. On the other hand, telephone interviews are less time-consuming and require no traveling effort. Thus, they were conducted if time restrictions or traveling distances did not allow face-to-face meetings. In these cases, the questionnaire was sent to the interviewee in advance via email. This helped a lot to avoid misunderstandings during the interview.

*4. Data analysis*: The interviews were transcribed literally and then analyzed using the constant comparison method. According to [17], this method consists of three steps: In *open coding*, particular themes or subjects that are of interests are marked in the text with codes or labels. After that, for every code, the marked passages in the interview texts are grouped together and read in their context in *axial coding*. Finally, hypotheses are generated from the data in selective coding. In this study, open and axial coding were used for analyzing the answers.

3) *Questionnaire*: The questionnaire consisted of three parts. The questions in the first part aimed at getting to know the background of the interviewee. The main questions were:
- What is your position?
- For how many years have you been involved in distributed development?

In the second part, we tried to get the background of the practitioner's company and its motivation and experiences in global software development. This part tried to answer research question 1. The main questions were:
- Why does your company do distributed development?
- What were the reasons for your company for establishing a development site?
- What are the experiences of your company in distributed software development?

Finally, we tried to reconstruct a concrete task allocation decision in order to answer research question 2. The interviewee was asked to consider one specific distributed development project and explain the task allocation. The main questions were:

- Please describe the project!
- What was your allocation decision?
- Why did you decide so?

*C. Execution*

1) *Data collection*: Interviews were conducted over a period of five months. Out of the twelve interviews included, nine were conducted in the US and three in India. The standardized questionnaire was used in eleven of the interviews. In the twelfth interview, the time available did not allow for using the complete questionnaire, so the interview was instead conducted by directly asking about the task assignment process.

Ten interviews were recorded and transcribed literally, while for the other two, detailed notes were taken during and after the interview.

2) *Data analysis*: The study was analyzed using QSR NVivo [16], a software tool for qualitative analysis and coding. Codes were created for all parts of the interview transcription. However, the major focus was on the second part of the questionnaire.

Most of the identified codes can be summarized into three categories:
- Project management: codes used were, for example, "Assignment Process", "Project Management", or "Requirements Engineering".
- Criteria for assignment: Codes used were, for example, "Costs", "Proximity to Client", or "Expertise".
- Problems: Codes were "Communication Problems", "Lack of Trust", and "Cultural Problems".

Other codes marked special issues such as "India" or "Granularity of the allocation".

## IV. RESULTS

The following section presents the results of the study. According to the order of the two research questions, the background of the participants' companies will be explained first. After that, the criteria that were applied for task allocation are shown for different types of distributed development, followed by a set of influencing factors and a comparison between the two results.

*A. Company Background*

1) *Reasons for doing distributed development*: The majority of the interviewed practitioners stated that in their companies, labor costs were one of the reasons for doing distributed development. Especially the initiation of global software development with sites or partners in India or China was usually strongly motivated by the expected cost savings.

Another major reason for initiating GSD was the availability of talented people worldwide. Nearly all of the interviewees mentioned accessing talent and people as a reason for doing distributed development. It was often mentioned that global development with India gave access to very large numbers of resources that could not be found at home. One interviewee pointed out:

> "[if] I need 100 people over the next three months, it is possible in India and maybe even in China but impossible in Germany or Finland."

Sometimes, talented people were scattered across different places and making them working together in a distributed way was easier than moving them all to one place.

Other reasons for initiating distributed development included: Knowledge about worldwide markets, being required by customers to work at different sites, and risk reduction by having knowledge distributed across several sites. In one case, distributed development was initiated due to the acquisition of foreign competitors.

2) *Site selection*: Some companies initiated distributed development by outsourcing work to other organizations. In this case, the locations of the available sites were dependent on the selected partner.

For the other organizations, the large companies had sites all over the world, often with large sites in far-Eastern low-cost countries. But also the smaller western companies had sites established in Asia: Of the two companies in the study with not more than 5 sites, one had development centers in India and China and the other one had a site in the Philippines.

TABLE 1. REASONS FOR INITIATING DISTRIBUTED DEVELOPMENT

| Cost | 9 |
|---|---|
| Access to people | 9 |
| Knowledge of markets | 1 |
| Required by customer | 1 |
| Risk reduction | 1 |
| Mergers and acquisitions | 1 |

TABLE II. CRITERIA FOR ESTABLISHING SITES

| Labor costs | 4 |
|---|---|
| Access to people | 3 |
| Proximity to customer | 2 |
| Historically driven | 3 |

The reasons for establishing a remote site mirrored the reasons for establishing GSD: Criteria for establishing a new site were the cost rate, the access to talents, or proximity to the customers. One interviewee, for instance, reported that his company often established new sites close to a university to get easy access to talented students. Another one said that her company (which developed custom software) always established a site near a customer as soon as the business with this customer grew beyond a certain size.

Sometimes the sites were established due to 'historical' reasons: Some practitioners reported that sites from acquired companies were continued or new sites were established because of prior collaborations with local partners.

3) *General Experiences*: Asked about their overall experiences with distributed software development, the interviewees responded quite pessimistically: A minority found the experiences rather positive, but most of them described them as "*lots of problems*" or "*serious difficulties*".

When the experiences were described as good, the interviewees usually meant that the problems of distributed development were not as big as expected. Only one person described a direct positive result of distributed collaboration:

*"Everyone, people are all involved all to the project and the style of decision-making is very collaborative."*

This shows that distributed development is usually seen by practitioners as a negative thing that is likely to produce a lot of problems – but, as one manager said:

*"We don't have a choice, we have to do it."*

The two main problems described were the impossibility of direct communication and differences in language and culture between the sites.

The impossibility of direct communication led to communication via email or telephone, which, as explained by several interviewees, was less effective. Another problem with indirect communication was finding the responsible person on the other side.

However, these problems seemed to be less significant when the distributed development was done on a small scale, assigning work to individuals instead of groups. Interviewees working within this type of distributed development complained much less about communication or coordination problems between the distributed individuals. One said:

*"[With] those individual interactions then the distributed part works because everybody is distributed. But when you have groups that are separated from each other they immediately build stovepipes and have difficulties."*

The reason for this difference could be that when work is done by distributed people instead of distributed groups, responsibilities are defined much clearer and there is also no possibility of developing rivalry or lack of trust between groups.

Differences in language and culture between sites were frequently mentioned by people doing global software development between different continents. These differences did not necessarily depend on the physical distances between countries. A manager of an Indian software company, for instance, stated:

*"We have a lot of problems in Asia, significantly less problems when it comes to working with the Anglo-Saxon work. So the US and Europe is easier to work with than if you work with Japanese or Chinese."*

Language problems even occurred, when all people involved spoke English. An Indian manager explained that Indians often worked as 'translators' between the English of people at Chinese sites and 'European' English.

But even in distributed development within one country, cultural problems can exist. One interviewee reported problems due to different working cultures at sites: Different quality standards led to misunderstandings and caused a lot of extra work.

Some practitioners mentioned further problems in control. From an organizational perspective, it is apparently hard to evaluate the working quality of remote sites. Therefore, one manager admitted that he had problems judging the performance of freshly-established sites and another manager reported that her company had a detailed process of judging all sites in order to cope with the problem of not knowing the sites.

### B. Criteria for Task Allocation

One of the first results of the interview analysis was that the interviewees' answers on task allocation within a specific project differed to a very large extent: They did not only describe different criteria for task allocation but also talked about very different ways of distributed development and task allocations. Therefore, in order to make the results comparable, the answers had to be grouped into several types.

The following section will thus first give a classification of distributed development and then name the identified criteria for task allocation for every group (research question 2).

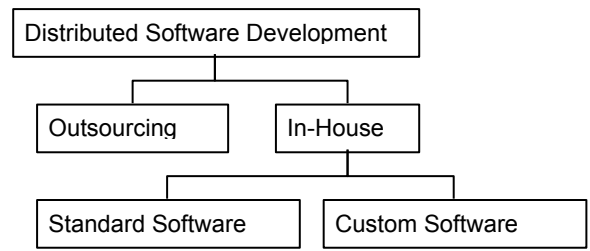

Figure 1. Types of distributed development

1) *A classification of distributed development*: In [14], a classification of distributed development was already introduced. The authors distinguish between on- and offshoring and between outsourcing and in-house development.

The interview results also revealed a major difference between outsourcing and in-house development. Between on- and offshoring, however, we found only minor differences in the way of distributing work. Instead, within in-house development, we found a considerable difference between organizations developing standard software products and organizations developing custom software.

Thus, the resulting types used in this analysis were:

**Outsourcing:** Parts or all of the software development is given to an outside contractor. From the client's perspective, task allocation is therefore rather a partner selection. Usually, after the activities have been assigned to a contractor, the client has little or no influence on the allocation of tasks to sites by the contractor. One interviewee described the control over the contractor's project management as:

*"This was like thrown over the fence and it was totally managed elsewhere".*

Usually, complete projects or large parts of a project are outsourced, whereas assignments in in-house development are typically on the task level.

**Standard software development**: An organization develops a software product that it sells to many different customers. Often, those organizations create a relatively large amount of proprietary knowledge and technology that is contained in specialized teams. This makes them very vulnerable to the effects of staff turnover (which is often experienced in India): One interviewee reported that in his company, it took new employees several years to become productive.

**Custom software development**: An organization develops individual software for clients. Compared to standard software development, the specialization and the creation of proprietary technology is lower. This makes the organizations less vulnerable to staff turnover: One interviewee working at a large organization with both a product development and service department reported that while the product development organization tried to keep the turnover rate as low as possible, the service department maintained a pool of available workforce in order to replace leaving staff.

Table 2 shows the distribution of the interviewed persons across the three types.

TABLE III. INTERVIEWED PERSONS PER TYPE

| Outsourcing | 3 |
|---|---|
| Standard Software | 5 |
| Custom Software | 4 |

2) *Outsourcing*: In two of the studied cases, costs were a major criterion for outsourcing and partner selection. In the third case, outsourcing was necessary due to previous downsizing and the focus in the selection was not so much on cost. Cost was not always seen as the pure development costs for a software development project. One interviewed manager particularly pointed out that the costs also included issues like taxes and had to be regarded over the complete lifecycle of the product.

Another important criterion was the *established relationship* with the contractor. One interviewee reported that her company had made a ten-year blanket contract with a software company that was chosen as the favorite contractor for all software development projects. Another manager explained why he favored an already established contractor:

*"[The contractor] has been building knowledge about our organization a long time. It has tremendous value; it means the risk of projects will go down because of that."*

Other criteria named were:

*Reliability of the contractor*: It was mentioned that reliability in the predictions and estimates of the contractor was often a criterion, which was sometimes more important than costs.

*Available expertise*: When outsourcing was initiated by the need for a certain expertise, that expertise was an important criterion.

*Proximity to users*: Even though one manager expressed the opinion that proximity to the customers or users was not of importance anymore, another reported that requirements engineering was assigned to the users' site while everything else was assigned to the contractor.

*Personal contacts*: One interviewee reported on a project in which the outsourcing partner was selected because of personal contacts of the manager in charge.

3) *Standard Software Development*: The most important criterion for task allocation was the expertise or skill set of the resources – this was named by all of the interviewees. Especially the larger companies seem to have specialized teams for different types of expertise. One manager reported:

*"We have database teams, we have runtime teams, we have security teams, we have language teams, UI teams, and some of those are distributed across different products."*

TABLE IV. CRITERIA IN OUTSOURCING

| Criterion | # Occurrences |
|---|---|
| Cost | 2 |
| Established relationship | 2 |
| Reliability | 1 |
| Expertise | 1 |
| Proximity to users | 1 |
| Personal contacts | 1 |

In that case, the specialization was so fixed that the scope of a new project was determined by the availability of the specialized team: In a new version of the product, the emphasis on the user interface, for example, was dependant on the availability of the user interface team. However, another interviewee reported a project in which some tasks were not assigned to the specialized team but to another site, because the specialized group was not available. In this project, fulfillment of the schedule had the highest priority.

Due to the specialized knowledge, the *turnover rate* also had an influence. While one manager reported that staff turnover was a reason for his company closing a subsidiary and transferring it to another country, others said that they tried to minimize staff turnover in order to avoid problems.

Another important criterion was the *proximity to markets*: Three interviewees reported that a required presence on emerging markets or knowledge about markets influenced their task allocation.

The *labor cost* rate at a site was sometimes a criterion: Two of the interviewees confirmed it as a criterion and for a third one, it was regarded at sites with low maturity. The other two denied that it had an impact.

TABLE V. CRITERIA IN STANDARD SOFTWARE DEVELOPMENT

| Criterion | # Occurrences |
|---|---|
| Expertise | 5 |
| Proximity to market | 3 |
| Labor costs | 3 |
| Turnover rate | 3 |
| Availability | 2 |
| Strategic planning | 2 |
| Maturity of site | 1 |
| Development quality | 1 |
| Personal trust | 1 |
| Product architecture | 1 |
| Time differences | 1 |
| Cultural differences | 1 |
| Willingness at site | 1 |

Other criteria named were:

*Availability*: While in one case, the availability of the specialized teams determined the scope of the development project, another manager described a case where work was assigned to a different team because of resources not being available at the specialized team. In this project, the focus was very much on finishing within schedule.

*Strategic planning*: Some managers reported that strategic considerations of transferring knowledge to certain sites influenced task allocation.

*Maturity of site*: The types of tasks that were assigned to a site depended on its maturity.

*Development quality*: The development quality and general capability of people at a site influenced the assignment of complex tasks.

*Personal trust*: To a certain degree, personal networks and trust between top management determined the organization of work.

*Product architecture*: Work packages were tried to be kept as independent as possible in order to minimize the communication needed between sites.

*Time difference*: One manager reported that testing was strategically placed at the geographical center of the sites in order to minimize the time difference to all other sites.

*Cultural differences*: In one case, complex tasks were not given to Asian sites because the US manager had the impression that their culture prevented them from actively trying to find solutions for new problems.

*Willingness at site*: In one company, the teams at the sites could decide themselves which tasks they wanted to do.

4) *Custom Software Development*: Similar to standard software development, the expertise of the workforce was one of the most important criteria for task allocation. However, the availability of people seems to be of equal importance. In one case, availability was considered even more crucial than expertise:

*"If we had been looking driven by the need for specific talent, we probably would not have looked in [the site we assigned to]. [...]But we didn't pick them because they had the best skill set; we picked them because they were available."*

TABLE VI. CRITERIA IN CUSTOM SOFTWARE DEVELOPMENT

| Criterion | # Occurrences |
|---|---|
| Expertise | 4 |
| Availability | 4 |
| Proximity to client | 4 |
| Labor cost | 2 |
| Strategic planning | 1 |
| Personal reasons | 1 |
| Political decisions | 1 |

The *cost rate* was also an important criterion for task assignment. Three of the interviews were conducted with persons working primarily in projects for US federal agencies, where they were forbidden to allocate work outside the US and thus could not leverage low cost rates. But two of those confirmed that in other projects within their company, cost rate per site was always a major criterion when it was possible to do work offshore.

Another major criterion was the required *proximity to the client*. All of the interviewees reported that certain tasks had to be done close to the customers' location, because they required either daily interaction with the customer (e.g., in requirements engineering) or access to the clients' machines (e.g., in integration).

Other criteria were:

*Strategic planning*: In one case, work was transferred to a specific site in order to strategically expand that site.

*Personal reasons*: One manager reported that one of the reasons for assigning work to a specific site was that a highly important project manager wanted to work from that location.

*Political reasons*: Sometimes, work had to be assigned to a site within the country of the (European) customer because for political reasons, the customer organization did not want to transfer jobs out of its home country.

## C. Further Findings

During the interviews, specific findings came up that had not been expected beforehand. Two of them will be presented in the following.

1) *The Cost Factor*: Expected cost savings is one of the most-cited benefits of GSD. Accordingly, cost savings was named as one of the main reasons for initiating distributed development. However, the cost rate was usually not a dominant criterion for task allocation in projects. One interviewee described the cost factor:

   *"I am sure it is in the back of someone's mind but it has never come up."*

One answer to this disproportion could be that the cost rate is rather a criterion for high-level, strategic decisions than for the actual allocation within projects: By expanding the low-cost sites, free resources and expertise are aggregated there, which automatically shifts future task allocation decision towards these sites. One manager, for instance, reported:

   *"I can see that there is more and more work in general coming to Bangalore and also to China. There are very big growth plans for these two areas."*

She also explained that her company regularly rotated Asian people to Western sites in order to build up expertise in the low-cost sites.

2) *Assigning work to India*: Nearly every interviewee had experiences with his organization working together with Indian companies, outsourcing work to India, or having a site established in India. Most of the work with India was initiated in order to achieve cost savings – only one practitioner mentioned the need for development around the clock as a reason for initiating work with India.

However, a series of problems were reported in distributed development with India: communication problems, inexperienced personnel, and high staff turnover.

Communication problems were reported due to the large distance to Western sites and because of cultural differences. Especially requirements engineering seems to be problematic. One interviewee described a project in which requirements engineering was to be done in India. However, it was not possible to effectively communicate from India with the users. In the end, people were sent from India to the US in order to gather requirements.

A lack of experience in India was also confirmed by an Indian manager:

   *"We get a lot of fresher's from the market but the retained people of experiences [are missing]."*

The problem of staff turnover was seen differently by the practitioners. Some saw it as a major problem that, together with the lack of experience, prohibited giving high-level work to India. One manager reported that his company had closed down a subsidiary in India because of the turnover rate and moved to the Philippines. However, to others the problem was not so severe. They explained it as a market problem caused by the high growth rates of software development in India. According to them, the problem could be handled by giving the Indian employees incentives to stay at their company.

## D. Limitations

The biggest limitation of the study is the size of the interview study. 12 interviews were conducted and for the analysis, the interviews were split up into three groups of three to five respondents each. This makes it impossible to make statements with any statistical significance and makes it hard to generalize the results.

The scope of the interview study is another limitation. Participants were selected based on personal contacts and availability. Thus, the sample is not representative. For example, three out of four participants talking about custom software development came from the same company. In addition, all three primarily had experiences with projects for US federal agencies.

Size and scope of the study make it questionable if the results cover the complete state of the practice in task allocation. However, we believe that the general classification and the phenomena described for each class can be generalized and reflect most of the GSD practices currently performed. Nevertheless, the study should be replicated in order to verify the results.

The terms used were sometimes not defined explicitly. For instance, when talking about "expertise", it was not always clear if this meant a specific technical expertise (e.g., for a certain programming language or an application domain) or a general capability. This created ambiguities.

Another limitation was the fact that from the practitioners' perspective, the interviews were given to an outsider. This might have influenced their answers, even though all of the interviewees seemed to respond very openly.

## V. CONCLUSION AND FUTURE WORK

This article presented an empirical study on criteria applied for task allocation in GSD. In interviews with practitioners in distributed development, it was revealed that most organizations initiated distributed development in order to access a larger pool of resources and leverage low labor costs. The results showed that several types of distributed development exist, differing largely in their process of allo-cating work. For the three types outsourcing, standard software, and custom software development, criteria applied in task allocation were identified.

Future work will have to further analyze task allocation. Since the relatively small number of participants makes the results not generalizable, a larger study on the practice of task allocation should be performed.

The study gave an insight to the criteria that currently *are* applied in task allocation. However, it did not address the question of what factors *should be* applied: High failure rates in distributed software development projects [6] imply that there is a need for improvement of project management in GSD, which potentially includes using a more comprehensive set of criteria for task allocation. Many of the interviewed practitioners could also report on experiences of projects having failed because the task allocation focused too much on a single criterion (often the cost rate), while other criteria (e.g., cultural differences, turnover rate) were neglected.

Therefore, future work should try to identify criteria that should be regarded systematically in task allocation in order to reduce the risks of global software development projects.

ACKNOWLEDGES

The authors would like to thank all participants of the interview study for giving their time and for providing insights into the practices of distributed software development. Professor Carolyn Seaman and Professor Peter Freeman provided valuable help and comments for the research presented here. Most of the work was done during a stay at the Fraunhofer Center for Experimental software Engineering, Maryland and was financially supported by the Otto A. Wipprecht Foundation. The authors also thank Sonnhild Namingha for proofreading the paper.

REFERENCES

[1] Bass, M., Paulish, D.: "Global Software Development Process Research at Siemens." *Third International Workshop on Global Software Development* ICSE 2004

[2] Carmel, E., Agarwal, R.: "Tactical Approaches for Alleviating Distance in Global Software Development." *IEEE Software* Vol. 18, No. 2, 2001

[3] Damian, D., Moitra, D.: "Global Software Development: How Far Have We Come?" *IEEE Software*, Vol. 23, No. 5, 2006

[4] Dubie, D.: "Outsourcing Moves Closer to Home." *CIO Today* December 18, 2007

[5] Edwards, H. K., Kim, J. H., Park, S., Al-Ani, B.: "Global Software Development: Project Decomposition and Task Allocation" *International Conference on Business and Information* (BAI2008)

[6] Fabriek, M., Brand, M. van de, Brinkkemper, S., Harmsen, F., Helms, R.W.,: "Reasons for Success and Failure in Offshore Software Development Projects." *European Conference on Information Systems*, 2008

[7] Grinter, R.E., Herbsleb, J.D., Perry, D.E.: "The geography of coordination: Dealing with Distance in R&D Work." *ACM Conference on Supporting Group Work 1999* (GROUP 99)

[8] A.T. Kearney Inc.: *Building the Optimal Global Footprint*, A.T. Kearney, Inc, Chicago, Illinois, USA, 2006.

[9] Krishna, S., Sahay, S., Walsham, G.: "Managing cross-cultural issues in Global Software Outsourcing." *Communications of the ACM* Vol. 47, No. 4, 2004

[10] Lamersdorf, A.: *Towards a global software development distribution model: Empirically-based model building for distributed software development*. Master Thesis, University of Kaiserslautern (November 2008). Retrievable at: http://wwwagse.informatik.uni-kl.de/staff/lamersdorf

[11] Mockus, A., Weiss, D. M.: "Globalization by Chunking: A Quantitative Approach". *IEEE Software* Vol. 18, No. 2, 2001

[12] Nachmias, D., Nachmias, C.: *Research Methods in the Social Sciences*. St Martin's Press, New York 1987

[13] neoIT: *Mapping Offshore Markets Update*. September 2005

[14] Prikladnicki, R., Audy, J. L. N., Damian, D., de Oliveira, T. C.: "Distributed Software Development: Practices and challenges in Different Business Strategies of Offshoring and Onshoring". *International Conference on Global Software Engineering* ICGSE 2007

[15] Prikladnicki, R., Audy, J. L. N., Evaristo, R.: "A Reference Model for Global Software Development: Findings from a Case Study." *International Conference on Global Software Engineering* ICGSE 2006

[16] QSR International: *NVivo 8 – Research Software for Analysis and Insight*. http://www.qsrinternational.com/products_nvivo.aspx

[17] Seaman, C.: "Qualitative Methods". In Shull, F. et al.: *Guide to Advanced Empirical Software Engineering*. Springer, 2008

[18] Setamanit, S., Wakeland, W.W., Raffo, D.: "Using simulation to evaluate global software development task allocation strategies." *Software Process: Improvement and Practice* 12(5): 491-503 (2007)

[19] Sooraj, P., Mohapatra, P.K.J.: "Developing an Inter-site Coordination Index for Global Software Development." *International Conference on Global Software Development*, ICGSE 2008

[20] Westner, M. K., Strahringer, S.: "Evaluation Criteria for Selecting Offshoring Candidates: An Analysis of Practices in German Businesses." *Journal of Information Technology Management* Vol. 19, No. 4, 2008